\documentclass[11pt,twocolumn]{article}
\usepackage[hmargin=0.75in, vmargin=0.75in]{geometry}
\usepackage[utf8]{inputenc} % for special symbols (umlaut, cedilla, etc.)
\usepackage[T1]{fontenc}

\usepackage[autostyle]{csquotes}\MakeOuterQuote{"} % fix quotation marks
\usepackage{mathtools, amssymb, amsthm}
\usepackage[backend=biber,style=phys,articletitle=true,biblabel=brackets,sorting=none]{biblatex}
\usepackage{verbatim} % type words like computer code (use \texttt{} for inline words
\usepackage{hyperref} 
\hypersetup{
    breaklinks,
    hidelinks,
    colorlinks=true, 
    linkcolor=black, % e.g., for equations
    filecolor=green, 
    citecolor=blue,       
    urlcolor=cyan
    }
\usepackage{calc} % for underlining section headings
\usepackage{titlesec} % customize sections
\usepackage{manyfoot} % for biblatex settings
\usepackage{relsize} % option to make math font larger
\usepackage{setspace} % adjusting line spacing
\usepackage{enumitem} % adjust indentation of lists
\usepackage[dvipsnames]{xcolor} % for adding color
\usepackage{xspace, xcolor, calligra} % variable customization (add ulem package for Vbar)
\usepackage{graphicx, float, placeins, subcaption} % for figures
\usepackage{booktabs, makecell, longtable} % for tables
\usepackage[hang,flushmargin,bottom]{footmisc} % fixes footnotes to bottom of page
\usepackage{ragged2e} % right align text
\usepackage{appendix} % for appendices
\usepackage{array}
\usepackage{etaremune} % reverse numbering of lists
\usepackage{etoolbox} % create link for email address
\usepackage[font=small,labelfont=bf,labelsep=period,tableposition=top]{caption} %format=hang
\usepackage{tikz} % for inlaying images
\usepackage{chemformula} % for chemical reaction symbols
\usepackage{widetext} % for making equations span both columns

\usepackage{newtxtext} % makes math look cooler; newtxmath

\usepackage{authblk}

\makeatletter
\renewcommand{\maketitle}{\bgroup\setlength{\parindent}{0pt}
\begin{center}
  \@title
\end{center}\egroup
}
\makeatother

\title{
    {\Large\textbf{Simulating spin--orbit coupled Bose--Einstein condensates: a concise overview for non-experts}}
    \vspace{1em} \\
    {\large Dylan R. Pollard\sups{1} \\
}}   
\DeclareMathAlphabet{\mathcalligra}{T1}{calligra}{m}{n}
\DeclareFontShape{T1}{calligra}{m}{n}{<->s*[2.2]callig15}{}

\newcommand*{\xhat}[1]{#1\kern-0.35em\hat{\phantom{#1}}}
\newcommand*{\xhatt}[1]{#1\kern-0.58em\hat{\phantom{#1}}}

\newdimen\LTcapwidth \LTcapwidth=6.6in

\newcommand{\sups}{\textsuperscript}

\newcommand\hl{%
  \bgroup\protected\def\sout{\bgroup \ULdepth =-1.2ex \ULset} % the 1.2 adjusts height
  \markoverwith{\textcolor{white}{\rule[-1ex]{1.1pt}{2.5ex}}} % the 0.1 adjusts width
  \ULon}

\newcommand*\oline[1]{%
  \vbox{%
    \hrule height 0.5pt%                  % Line above with certain width
    \kern0.25ex%                          % Distance between line and content
    \hbox{%
      \kern-0.12em%                        % Distance between content and left side of box, negative values for lines shorter than content
      \ifmmode#1\else\ensuremath{#1}\fi%  % The content, typeset in dependence of mode
      \kern-0.12em%                        % Distance between content and left side of box, negative values for lines shorter than content
    }}}

\makeatletter
\newcommand\myemail[3]{\edef\@tempa{mailto:#1?subject=#2 }%
\edef\@tempb{\expandafter\html@spaces\@tempa\@empty}%
\href{\@tempb}{#3}}
\catcode\%=11
\def\html@spaces#1 #2{#1%20\ifx#2\@empty\else\expandafter\html@spaces\fi#2}
\catcode\%=14
\makeatother

\newcommand*{\belowrulecolor}[1]{% 
  \noalign{% 
    \kern-\belowrulesep 
    \begingroup 
      \color{#1}% 
      \hrule height\belowrulesep 
    \endgroup 
  }
}

\newcommand*{\aboverulecolor}[1]{% 
  \noalign{% 
    \begingroup 
      \color{#1}% 
      \hrule height\aboverulesep 
    \endgroup 
    \kern-\aboverulesep 
  }%
}

\newcommand{\micron}{$\mu$m}
\newcommand{\angst}{\AA}
\newcommand{\wavenumber}{cm$^{-1}$}

\newcommand{\degc}{$^\circ$C}
\newcommand{\degf}{$^\circ$F}
\newcommand{\degr}{$^\circ$R}
\newcommand{\degs}{$^\circ$}

\newcommand{\primary}{{\sf 1\/}$^\circ$}
\newcommand{\secondary}{{\sf 2\/}$^\circ$}
\newcommand{\tertiary}{{\sf 3\/}$^\circ$}
\newcommand{\quartenary}{{\sf 4\/}$^\circ$}

\newcommand{\uv}[1]{\mbox{\boldmath $\delta$}_#1}
\newcommand{\pd}[1]{\frac{\partial}{\partial #1}}
\newcommand{\pdd}[2]{\frac{\partial #1}{\partial #2}}
\newcommand{\secondpd}[1]{\frac{\partial^2}{\partial #1^2}}
\newcommand{\secondpdd}[2]{\frac{\partial^2 #1}{\partial #2^2}}
\newcommand{\cross}{\mbox{\boldmath $\times$}}
\newcommand{\vdot}{\mbox{\boldmath $\cdot$}}
\newcommand{\bv}[1]{\mathbf #1}
\newcommand{\st}{\mbox{\boldmath $\tau$\unboldmath}}
\newcommand{\pit}{\mbox{\boldmath $\pi$\unboldmath}}
\newcommand{\idt}{\mbox{\boldmath $\delta$\unboldmath}}
\newcommand{\altt}{\mbox{\boldmath $\epsilon$\unboldmath}}
\newcommand{\dd}[1]{\frac{d}{d #1}}
\newcommand{\ddd}[2]{\frac{d #1 }{d #2}}
\newcommand{\seconddd}[1]{\frac{d^2}{d #1^2}}
\newcommand{\secondddd}[2]{\frac{d^2 #1 }{d #2^2}}
\newcommand{\thirddd}[1]{\frac{d^3}{d #1^3}}
\newcommand{\thirdddd}[2]{\frac{d^3 #1 }{d #2^3}}
\newcommand{\fourthdd}[1]{\frac{d^4}{d #1^4}}
\newcommand{\fourthddd}[2]{\frac{d^4 #1 }{d #2^4}}
\newcommand{\sd}[1]{\frac{D #1 }{D t}}

\newcommand{\sci}[2]{#1$\times$10$^{#2}$}

\newcommand{\matlabr}{\textsc{Matlab}\raise 1ex\hbox{\tiny \textregistered}\xspace}
\newcommand{\matlab}{\textsc{Matlab}\xspace}
\newcommand{\me}{Dylan R. Pollard\xspace}
\bibliography{main}

\captionsetup[table]{singlelinecheck=off} % left-align table captions
\emergencystretch 6pt % ensures that all lines stay within margins
\renewcommand*{\thepage}{- \arabic{page} -} % dashes with page numbers
\usepackage{enumitem}\setlist{nolistsep} % get rid of spaces around lists
\titleformat{\section}{\normalfont\fontsize{12}{11}\bfseries\filcenter}{\thesection}{1em}{} % adjust section heading font size and make centered
\renewcommand{\thesection}{\Roman{section}.\hspace{-0em}}

\titlespacing\section{0pt}{12pt plus 4pt minus 2pt}{0pt plus 2pt minus 2pt}
\renewcommand{\abstractname}{} % hides the word "abstract"
\renewenvironment{abstract}
 {\small
  \begin{center}
  \bfseries \abstractname\vspace{-.5em}\vspace{0pt}
  \end{center}
  \list{}{%
    \setlength{\leftmargin}{18mm}% <---------- CHANGE HERE
    \setlength{\rightmargin}{\leftmargin}%
  }%
  \item\relax}
 {\endlist}

\newcommand{\PRLsep}{\vspace{0.25em}\noindent\makebox[\linewidth]{\resizebox{0.5\linewidth}{1.5pt}{$\bullet$}}\vspace{0.5em}}

\begin{document}
\parskip = 6pt % adjust spacing between paragraphs
\setstretch{1} % line spacing
\raggedbottom

%%%%%%%%%%%%%%%%%%%%%%%%%%%%%%%%%%%%%%%%%%%%%%%%%%%%%%%%%%%%%%
%\thispagestyle{empty}

\twocolumn[
\begin{@twocolumnfalse}
    \maketitle
    \vspace{-3em}
    
    \begin{abstract}
		Cold ensembles of bosons are a useful platform for studying many-body quantum states present in quantum technologies, and simulations of these systems are convenient for streamlining design of such technologies. This paper provides an overview of spin--orbit coupled Bose--Einstein condensates (SOC BECs), along with an overview of simulation methods and a case study on a novel "microemulsion" phase. These SOC-BECs are further described in the context of coherent states quantum field theory, utilizing the complex Langevin sampling scheme as a numerical algorithm. This approach for studying SOC BECs is a versatile and powerful alternative to particle-based methods and has promise for simulating other bosonic quantum states.
 	\end{abstract}
	
	\vspace{2em}
\end{@twocolumnfalse}
]

%%%%%%%%%%%%%%%%%%%%%%%%%%%%%%%%%%%%%%%%%%%%%%%%%%%%%%%%%%%%%%%%%%%%%%%%%%%%%%%%%%%%%%%%%%%%%%

\section{INTRODUCTION}\label{intro}

\let\thefootnote\relax\footnotetext{\sups{1} \texttt{pollard@ucsb.edu}}

Many proposed quantum technologies (e.g., cold-atom quantum computing) leverage a variety of exotic quantum states, such as superfluidity, superconductivity, Bose--Einstein condensation, quantum disordered spin states, etc. To accelerate the design of quantum technologies with such exotic quantum states, we can perform simulations to understand the underlying phenomena, optimize system parameters, and predict emergent behaviors under different conditions. Cold bosons are one platform for studying these quantum states. These ultra-cold (nK scale) bosonic systems can be formed via different cooling and confining techniques until a large fraction of atoms occupies a single-particle state, inducing spontaneous symmetry breaking into a Bose--Einstein condensate (BEC) \cite{NegeleQMPS1998}. These bosonic ensembles, when isolated and localized, allow for probing of their quantum many-body physics to high precision. However, for many boson systems that have high density or are subject to artificial gauge fields, conventional particle-based Monte Carlo simulations are expensive and can be numerically unstable, whereas field-theoretic simulations (FTS) are ideal, particularly via coherent states (CS) quantum field theory \cite{NegeleQMPS1998,DelaneyCLsamp2020}. CS field theory, when coupled with the complex Langevin sampling technique, allows for simulation of dense BECs under artificial gauge fields \cite{DelaneyCLsamp2020}. One example is an artificial spin--orbit coupling (SOC), which intertwines a particle's intrinsic spin and momenta. SOC is naturally pervasive in material systems such as topological insulators but does not occur naturally in neutral atom BECs \cite{GoldmanSOCrev2014}. By imbuing ultra-cold bosons with SOC, one can access novel phases of SOC BECs, such as ringed, hexagonal, striped, crystalline, and, recently, a "microemulsion" phase \cite{McGarrigleMicro2023}. In this Letter, we will discuss the quantum many-body physics and simulation techniques for SOC-BECs and how one can observe a microemulsion phase under certain conditions.

\section{PARTICLES TO FIELDS}\label{background}

The need to impose symmetry properties complicates the coordinate representation of quantum many-body systems at finite temperature. Under boson statistics, wavefunctions are symmetric with respect to the exchange of identical particles. Let us consider a collection of $n$ spinless, indistinguishable quantum particles at temperature $T$ and volume $V$. The partition function for such a collection of bosons can be written as
\begin{align}\label{partitionFun}
\mathcal{Z}_B &= \frac{1}{n!}\sum_P \int dR \rho_D(R,PR';\beta) \nonumber \\
  & \equiv \frac{1}{n!}\sum_P \ \mathcal{Z}_D(n,V,T)
\end{align}
where the partition function $\mathcal{Z}_B$ is the sum of all permutations of the $n$ bosonic particles and imbeds proper Bose symmetry. Here, $\rho_D(R,R';\beta) = \langle R | \mathrm{exp}(-\beta \hat{H}) | R' \rangle$ is the \textit{equilibrium density matrix}, where $|R\rangle$ is a many-body position state that contains all $n$ atomic coordinates; $\hat{H}$ is the many-body Hamiltonian operator; and $\beta = 1/k_BT$. The trace of $\rho_D$ yields the partition function for distinguishable particles, $\mathcal{Z}_D$, and the operator $P$ denotes a permutation of interchanging particles $R$ and $R'$. One can then perform the routine steps of Trotterizing the density matrix and taking the limit of continuous imaginary time to get the continuous (Feynman) imaginary-time path integral for $n$ distinguishable bosons
\begin{equation}
	\mathcal{Z}_D(n,V,T) = \prod_{\alpha = 1}^n \left[ \int \mathcal{D}\mathbf{r}_\alpha \right] \mathrm{exp} \big( -S[\{\mathbf{r}_\alpha\}] \big)
\end{equation}
where $S[\{\mathbf{r}_\alpha\}]$ is the action and given in Ref. \cite{GlennFTS2023}, and $\mathbf{r}_\alpha$ represents the coordinate vector for the $\alpha^\mathrm{th}$ particle. Intuitively, Feynman's path integral approach amounts to Brownian diffusive motion of particles with imaginary-time period $\beta$. At low temperature and high density, the quantum fluid is attempting to lower its kinetic energy, and as $T \rightarrow 0$ ($\beta \rightarrow \infty$), all permutations contributing to $\mathcal{Z}_B$ are equally likely \cite{GlennFTS2023}. These permutations give rise to a wide variety of quantum phenomena, such as the spin liquid behavior of frustrated magnets \cite{balentsSpin2010} and superfluidity of liquid helium \cite{CeperleyHelium1995}, due to the \textit{exchange interactions} at low temperatures. For example, even gaseous systems of bosons can form a BEC, defined as a bosonic system where the zero-momentum ground state is macroscopically occupied. This unique phase transition is different from its classical analog in that condensation occurs in momentum space instead of coordinate space with an order parameter of a complex phase instead of fluid density.

\begin{figure}
\centerline{\includegraphics[scale=0.15]{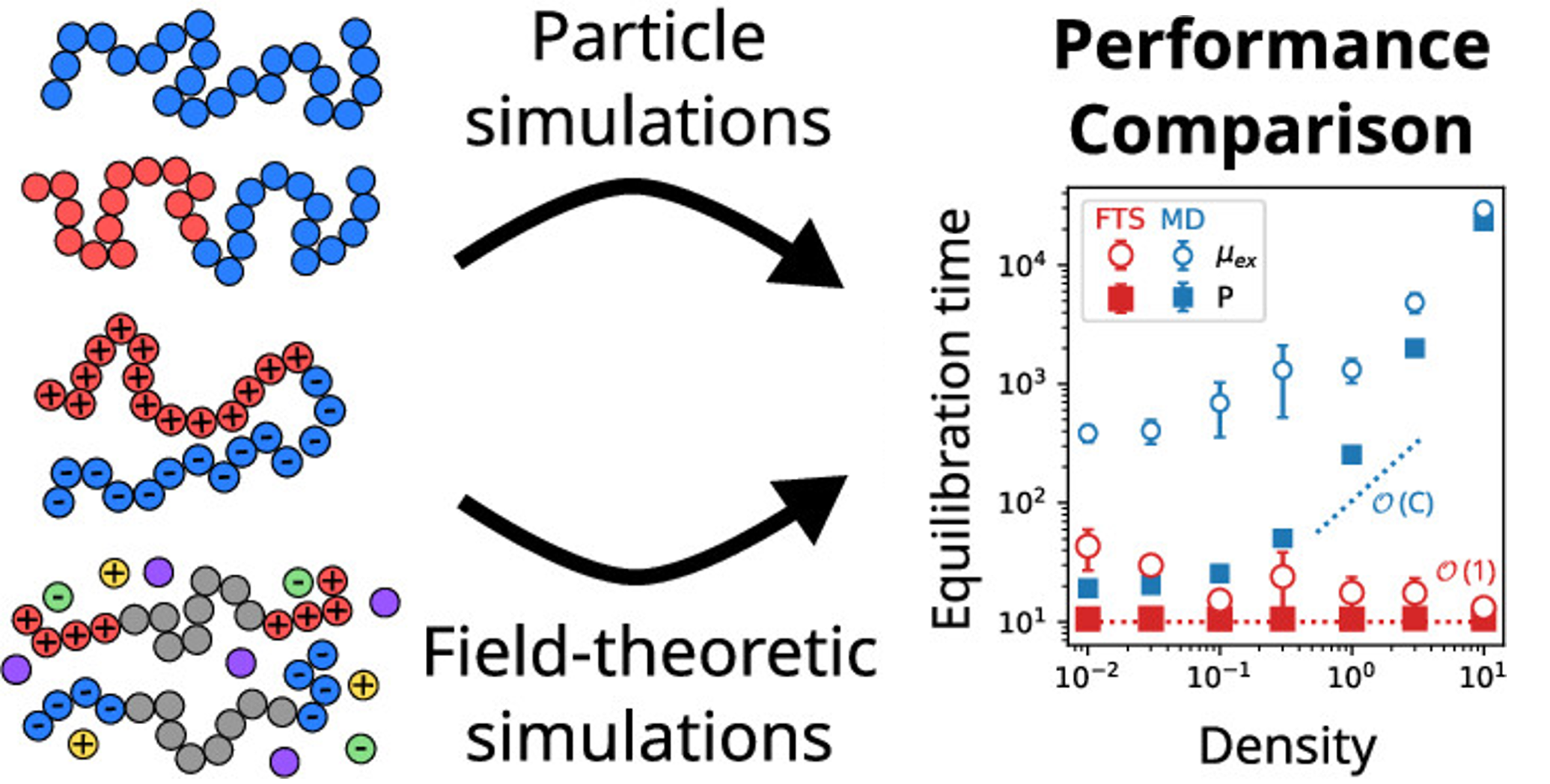}}
\caption{Computation cost for field-theoretic simulations and molecular dynamics (e.g., PIMC) versus particle density. Reproduced from Ref. \cite{LequieuFTSMD2024} by J. Lequieu (CC-BY 4.0).}
\label{computeCost}
\end{figure}

For many years, \textit{path integral Monte Carlo} (PIMC) was the only simulation method that yielded numerically accurate estimates of continuum many-body systems of bosons at finite temperature \cite{GlennFTS2023}. PIMC has been used to study liquid $^\mathrm{4}$He \cite{CeperleyHelium1995,BoninsegniWorm2006}, quantum crystalline phases \cite{GopalSOCcryst2013}, and ultra-cold gases \cite{PolletColdGas2012}, among many other bosonic systems. However, for systems over a few thousand particles, PIMC remains expensive due to the multitude of quantum permutations (see Figure \ref{computeCost}). Because of quantum correlations, a trial move in PIMC involves sequentially iterating through imaginary time slices over each individual particle, generating an overall diffusive motion but resulting in high computational cost as system size increases. Designing efficient permutation sampling schemes is also challenging due to system-dependent interactions, such as the harshly repulsive potentials of liquid helium; thus, the prospective for future improvements is unclear.

Instead of the particle description of PIMC, \textit{coherent states (CS) field theory} offers a field-theoretic simulations (FTS) method of representing the imaginary time path integral of quantum statistical mechanics. Since Bose statistics are embedded within CS field theory, there is no need for summing over the permutations in Equation \ref{partitionFun}. For many-body quantum systems, quantum field theories are developed in an abstract occupation-number basis, which represents the number of particles in a complete, orthogonal set of one-particle quantum states, rather than in the particle coordinate basis. This second quantization framework inherently incorporates Bose or Fermi symmetry through commutation relations of raising and lowering operators in the Hamiltonian. The partition function in this basis can be reformulated as an imaginary-time path integral by inserting complete sets of coherent states---linear combinations of occupation-number states---along the trajectory \cite{NegeleQMPS1998}. This leads to a CS field theory with complex fields defined in $d+1$ dimensions, where the extra dimension corresponds to imaginary time. 

In other words, combining second quantization with Feynman's path integral formalism yields a framework amenable to field-theoretic simulations. Indexing a single particle state by plane wave mode $\mathbf{k}$, a coherent state can be defined as
\begin{equation}
	|\phi_\mathbf{k} \rangle \equiv \sum_{n_\mathbf{k} = 0}^\infty \frac{1}{\sqrt{n_\mathbf{k}!}}\phi_\mathbf{k}^{n_\mathbf{k}} | n_\mathbf{k} \rangle
\end{equation}
where $|n_\mathbf{k} \rangle$ is an occupation number state, and $\phi_\mathbf{k}$ is a complex number. Note that the set of coherent states is not orthogonal since they overlap in momentum space. To construct a coherent states path integral, recall that the algebra of second quantization relies on varying particle number, so it is natural to consider the grand canonical partition function $\mathcal{Z}_G(\mu,V,T)$, where $\mu$ is the chemical potential. Using the grand canonical density matrix and second quantization relations \cite{NegeleQMPS1998}, $\mathcal{Z}_G$ in terms of coherent states is given by
\begin{equation}\label{Zbec}
	\mathcal{Z}_G(\mu,V,T) = \int \mathcal{D}(\phi^*,\phi) \ \mathrm{exp}\big(-S[\phi^*,\phi]\big)
\end{equation}
where $\mathcal{D}(\phi^*,\phi)$ is shorthand for the product over momenta $\mathbf{k}$ of the real and imaginary components of $d\phi_\mathbf{k}$. Depending upon the particular bosonic system, the action $S[\phi^*,\phi]$ can have multiple terms which are derived elsewhere (see Ref. \cite{GoldmanSOCrev2014}). In the coherent states basis, a 3-dimensional assembly of $n$ bosons confined with an external potential and arbitrary gauge fields in typical experiments assumes an action \cite{GoldmanSOCrev2014,GlennFTS2023} of

\newpage
\begin{widetext}
  \begin{align}\label{CSaction}
    S[\phi^*,\phi] &= \int_0^\beta d\tau \int_V d^3r \phi^*(\mathbf{r},\tau+) \left[ \pd{\tau} - \mu + U_\mathrm{ex}(\mathbf{r}) \right] \phi (\mathbf{r},\tau) \nonumber \\ 
    &+ \int_0^\beta d\tau \int_V d^3r \phi^*(\mathbf{r},\tau+) \left( \frac{1}{2m}\big[\hat{\mathbf{p}} - \boldsymbol{\mathcal{A}}(\mathbf{r})\big]^2 \right)\phi(\mathbf{r},\tau) \nonumber \\
    &+ \frac{g}{2} \int_0^\beta d\tau \int_V d^3r \big[ \phi^*(\mathbf{r},\tau+)\phi(\mathbf{r},\tau) \big]^2
  \end{align}
\end{widetext}

\noindent
where $\hat{\mathbf{p}} = -i\hbar\nabla$ is the canonical momentum operator; $U_\mathrm{ex}(\mathbf{r})$ is an external trapping potential; $\boldsymbol{\mathcal{A}}(\mathbf{r})$ is an \textit{artificial (vector) gauge potential}; and $g$ is the contact repulsion strength. The gauge field $\boldsymbol{\mathcal{A}}(\mathbf{r})$ is of interest in that it can be Abelian (\textit{e.g.}, externally imposed rotational flow) or non-Abelian (\textit{e.g.}, containing Pauli matrices). In the latter, this type of gauge field is evident in spin--orbit coupling, as discussed in Section \ref{soc}

\section{SPIN--ORBIT COUPLING}\label{soc}

\begin{figure}[H]
\centerline{\includegraphics[scale=0.3]{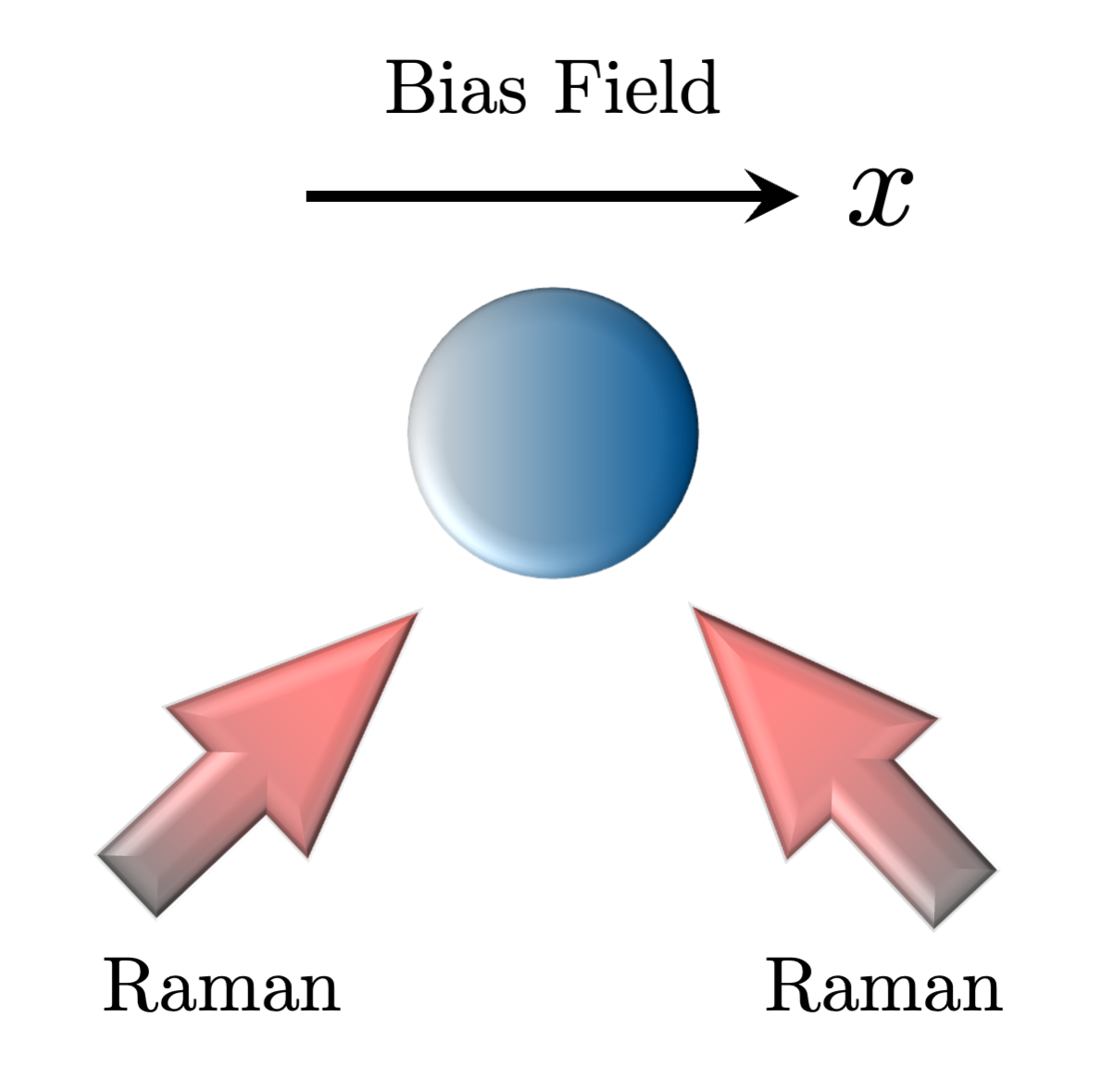}}
\caption{Example experimental realization of artificially induced SOC. Two Raman lasers intersect the cloud with vertex orthogonal to B-field direction.}
\label{SOCBECexp}
\end{figure}

To probe the many-body physics of ensembles of particles with relatively high precision, one can impose external potentials and artificial gauge fields to cool, isolate, and localize particles to form dense, interacting systems such as BECs. Through methods such as laser cooling, evaporative cooling, or magnetic/optical trapping, even dilute gases of bosons can be tuned to be highly interacting and correlated, allowing for observance of a variety of exotic quantum states. One particular phenomenon, called \textit{spin--orbit coupling} (SOC), occurs when a system of multi-level bosons are cooled and confined within spatially varying laser fields that split the ground state into different energy levels and couple different levels together \cite{StanescuSOCtheory2008}. The canonical example of this is a degenerate cloud of \sups{87}Rb atoms in a crossed optical dipole trap with a magnetic field bias (see Figure \ref{SOCBECexp}). The two Raman laser couple the states $| F = 1, \, m_f = -1 \rangle$ and $| F = 1, \, m_f = 0 \rangle$ (higher splitting due to B-field; out of resonance), and states $| F = 1, \, m_f = 0 \rangle$ and $| F = 1, \, m_f = 1 \rangle$ (remains in resonance) \cite{ZhangSOCexpRev2016}. This splitting results in a pseudo-spin-1/2 system, where $F$ represents the total hyperfine angular momentum, and $m_f$ represents the corresponding quantum number.

\noindent
\textit{2D Isotropic Rashba SOC}

Bose--Einstein condensation allows various macroscopic quantum states depending upon the external potential $U_\mathrm{ex}(\mathbf{r})$, the interaction potential between particles $V(\mathbf{r})$, and the presence of SOC due to $\boldsymbol{\mathcal{A}}(\mathbf{r})$. One simple example of SOC corresponds to 2D isotropic Rashba SOC \cite{BychkovOscill1984} with vector gauge potential
\begin{equation}\label{RashbaGauge}
	\boldsymbol{\mathcal{A}} = \hbar\kappa (\boldsymbol{\sigma}^x \mathbf{e}_x + \boldsymbol{\sigma}^y \mathbf{e}_y )
\end{equation}
where $\boldsymbol{\sigma}^\mu$ are Pauli spin matrices, and $\kappa$ is a parameter for SOC strength. For this paper, "SOC BEC" refers to a BEC with 2D isotropic Rashba SOC. 

Different SOC BEC states exist depending upon if the state is interacting ($V(\mathbf{r}) \neq 0$) or if the state is homogeneous ($U_\mathrm{ex}(\mathbf{r}) = 0$). For example, an interacting homogeneous SOC BEC has either a "plane wave" phase or a "striped superfluid" phase depending on its interaction strength \cite{GoldmanSOCrev2014}. On the other hand, the phase of a \textit{non}interacting homogeneous SOC BEC has an energy dispersion of a "ring-bottom" paraboloid with a continuum minima (ring) of $k_\mathrm{min}$ values, namely a Rashba circle. Moreover, an SOC BEC with harmonic $U_\mathrm{ex}$ and increasingly strong interactions, yields a variety of phases, including ringed, hexagonal lattice, and striped phases \cite{SinhaTrapped2011}.

\vspace{0.5em}
\noindent
\textit{Numerical Simulations of SOC BECs}

The applicability of the chosen simulation scheme, commonly either (PI)MC or FTS, depends upon the positive definitiveness of the particle distribution and is due to a "sign problem". The sign problem is a result of the non-positive-definiteness of the representative statistical weight for the particle distribution. It is present for complex and extensive Hamiltonians for FTS (both classical and quantum) \textit{and} PIMC for SOC BECs, resulting in complex oscillations that thwart numerical convergence.

To circumvent the sign problem, complex Langevin (CL) dynamics proves to be a viable solution by adaptively sampling along nearly stationary phase trajectories, and if trajectories converge to a stationary solution and obey the CL correctness criteria, they are proven free of any bias \cite{LeeConvergence1994,DelaneyCLsamp2020}.The CL algorithm computes expectation values via operator averages over Langevin time. A stable algorithm leverages an off-diagonal descent scheme that effectively decouples $\phi_{\alpha,j}^*(\mathbf{r},t)$ and $\phi_{\alpha,j}(\mathbf{r},t)$ to linear order, resulting in the following CL equations of \mbox{motion \cite{DelaneyCLsamp2020}}
\begin{align}
	\pd{t}\phi_{\alpha,j}(\mathbf{r},t) = -\frac{\delta S[\phi,\phi^*]}{\delta \phi_{\alpha,j}^*(\mathbf{r},t)} + \eta_{\alpha,j}(\mathbf{r},t) \nonumber \\
	\pd{t}\phi_{\alpha,j}^*(\mathbf{r},t) = -\frac{\delta S[\phi,\phi^*]}{\delta \phi_{\alpha,j}(\mathbf{r},t)} + \eta_{\alpha,j}^*(\mathbf{r},t)
\end{align}
where $\eta_{\alpha,j}(\mathbf{r},t)$ and $\eta_{\alpha,j}^*(\mathbf{r},t)$ are complex-conjugate white noise terms with zero mean and variance that ensures the noise precisely captures both quantum and thermal fluctuations \cite{McGarrigleMicro2023}, and periodic boundary conditions are implemented in the spatial and imaginary-time domains.

\vspace{0.5em}
\noindent
\textit{Spin Microemulsion Phase}

One case study for understanding SOC BECs is the proposed spin microemulsion phase which has been shown to have undulating pseudo-spin domains with energy dispersions that follow the Rashba circle \cite{McGarrigleMicro2023}. The assembly of bosons in this particular phase are well-described by the Hamiltonian which, in terms of field operators, is given by
\begin{align}
	H = \sum_{\alpha\gamma} \int d^2r \ \hat{\psi}_\alpha^\dag (\mathbf{r}) \left[ \frac{1}{2m} ( \hat{\mathbf{p}}\mathbf{I} - \boldsymbol{\mathcal{A}} )^2 - \mu\mathbf{I} \right]_{\alpha\gamma} \hat{\psi}_\gamma (\mathbf{r}) \nonumber \\
	\hspace{0em}+ \frac{1}{2}\sum_{\alpha\gamma} \int d^2r \ \hat{\psi}_\alpha^\dag (\mathbf{r}) \hat{\psi}_\gamma^\dag (\mathbf{r}) ( g_0\mathbf{I} + g_1\boldsymbol{\sigma}^x )_{\alpha\gamma} \hat{\psi}_\alpha (\mathbf{r}) \hat{\psi}_\gamma (\mathbf{r})
\end{align}
Here, the Hamiltonian is normal ordered, and the repulsive coupling constants $g_0$ and $g_1$ correspond to like and unlike pseudo-spin scattering events and are elements of a symmetric coupling constants matrix \cite{McGarrigleMicro2023}. For studying the universal behavior of SOC BEC phases, we can define parameters and nondimensional groups. Let $\ell \equiv \sqrt{\hbar^2/(2m\mu_\mathrm{eff})}$ be a natural length scale and $\mu_\mathrm{eff} \equiv \mu - \hbar^2\kappa^2/(2m)$ be an effective chemical potential. Define the dimensionless repulsion scale $\tilde{g} \equiv 2mg_0/\hbar^2$, SOC strength $\tilde{\kappa} \equiv \kappa \ell$, temperature $\tilde{T} \equiv 1/(\beta\mu_\mathrm{eff})$, and miscibility parameter $\eta_g \equiv g_1/g_0$. 

To access finite temperature, we can combine these parameters and dimensionless groups with Equations \ref{CSaction} and \ref{RashbaGauge}. After discretizing Equation \ref{CSaction} in imaginary time (necessary for numerical simulation) and summing over pseudo-spin species $\alpha$, one gets the coherent state-dependent action
\begin{align}
	S[\phi,\phi^*] &= \sum_\alpha \sum_{j=0}^{N_\tau - 1} \int \hspace{-0.2em} d^2r  \hspace{0.2em} \phi_{\alpha,j}^* (\mathbf{r}) \big[ \phi_{\alpha,j} (\mathbf{r}) - \phi_{\alpha,j-1} (\mathbf{r}) \big] \nonumber \\
	&+ \frac{\tilde{\beta}}{N_\tau} \sum_{\alpha,\gamma} \sum_{j=0}^{N_\tau - 1} \int \hspace{-0.2em} d^2r  \hspace{0.2em} \phi_{\alpha,j}^* (\mathbf{r}) \hat{\mathcal{K}}_{\alpha\gamma} \phi_{\gamma,j-1} (\mathbf{r}) \nonumber \\
	&+ \frac{\tilde{\beta}\tilde{g}}{2N_\tau} \sum_{\alpha,\gamma} \sum_{j=0}^{N_\tau - 1} \int \hspace{-0.2em} d^2r  \hspace{0.2em} \phi_{\alpha,j}^* (\mathbf{r}) \nonumber \\
	&\times \phi_{\gamma,j}^* (\mathbf{r}) \big( \mathbf{I} + \eta_g\boldsymbol{\sigma}^x \big)_{\alpha\gamma} \phi_{\gamma,j-1}(\mathbf{r}) \phi_{\alpha,j-1}(\mathbf{r})
\end{align}
Here, imaginary time is discretized into $j$ slices for \mbox{$\tau \in [0,\tilde{\beta}]$}, where $\tilde{\beta} = 1/\tilde{T}$, and the matrix $\hat{\mathcal{K}}_{\alpha\gamma}$ is
\begin{equation}
	\begin{pmatrix}
		-\tilde{\nabla}^2 - 1 & -2\tilde{\kappa}[ -i\partial_{\tilde{x}} - \eta_\mathrm{SOC}\partial_{\tilde{y}} ] \\
		-i\partial_{\tilde{x}} + \eta_\mathrm{SOC}\partial_{\tilde{y}} & -\tilde{\nabla}^2 - 1
	\end{pmatrix}
\end{equation}
where $\eta_\mathrm{SOC} = 1$ since an isotropic SOC is assumed. Thus, there are three space-imaginary time dimensions in which the complex coherent states fields are discretized and primed for numerical simulation.

\begin{figure}[H]
\centerline{\hspace{0.7em}\includegraphics[scale=0.175]{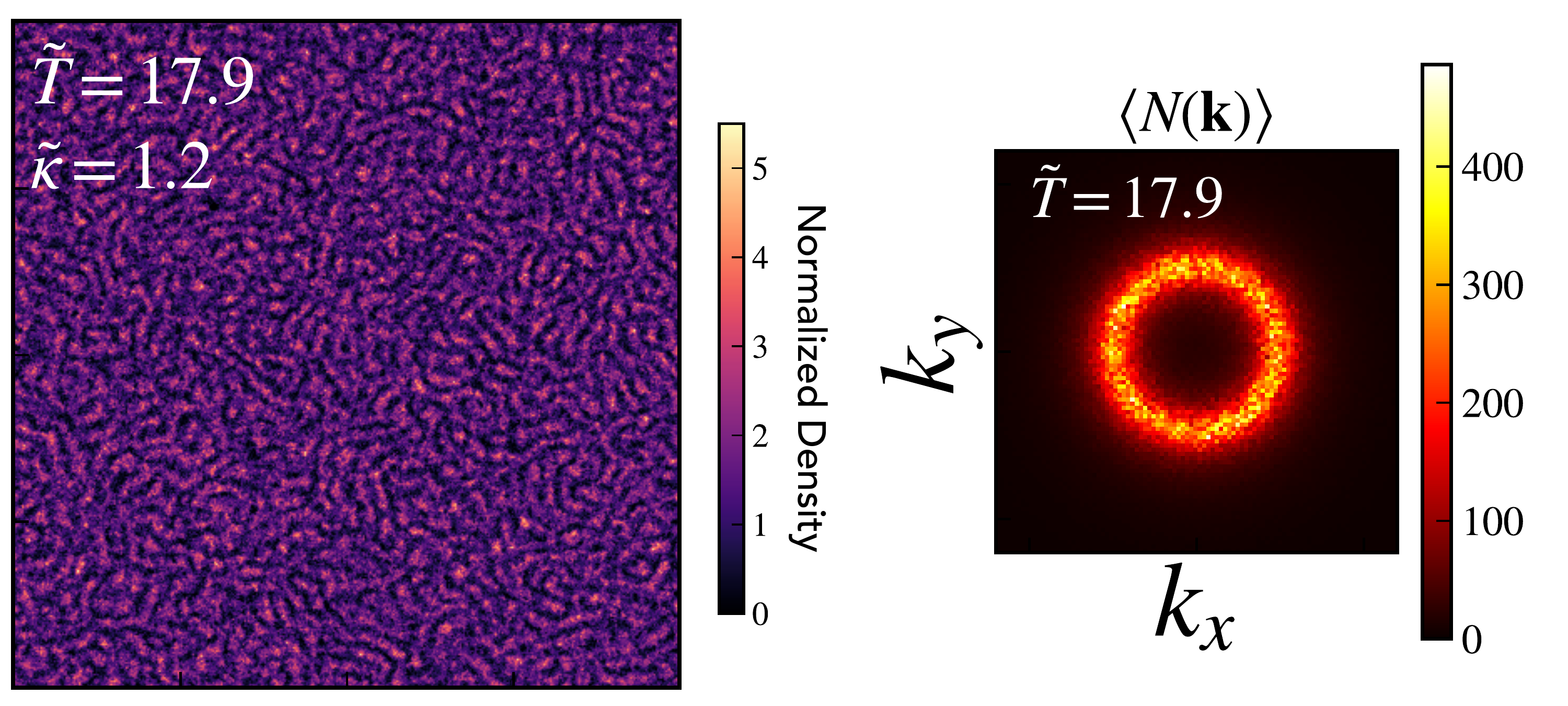}}
\caption{Normalized density profile and thermally averaged momentum distribution of pseudo-spin bosons in the $|\hspace{-0.3em}\uparrow\rangle$ basis state. Here, $\tilde{g} = 0.05$, and $\eta_g = 1.1$. Adapted with permission from McGarrigle \textit{et al.} \cite{McGarrigleMicro2023}.}
\label{microemulsion}
\end{figure}

Implementing the CL sampling scheme as described in the previous section, one can now run the algorithm and compute quantities such as density profiles $\rho_\alpha[\phi,\phi^*;\mathbf{r}] = 1/N_\tau \sum_{j=0}^{N_\tau-1}\phi_{\alpha,j}^*(\mathbf{r})\phi_{\alpha,j-1}(\mathbf{r})$, the momentum distribution $N[\phi,\phi^*;\mathbf{r}] = A/N_\tau \sum_\alpha \sum_{j=0}^{N_\tau-1} \tilde{\phi}_{\alpha,j,-\mathbf{k}}^*\tilde{\phi}_{\alpha,j-1,\mathbf{k}}$ (where "$^\sim$" signifies discrete Fourier transform), and other thermodynamic quantities. The density profile and momentum distribution for the spin microemulsion SOC BEC was determined in Ref. \cite{McGarrigleMicro2023} and is reproduced with permission in Figure \ref{microemulsion}. This unique phase embodies an isotropic "spin emulsion" structure, has a characteristic domain width of $\pi/(2\tilde{\kappa})$, and exhibits no long-range order. Additionally, the most occupied momentum state is that with wavevector $|\mathbf{k}| = \tilde{\kappa}$ which is that of canonical 2D isotropic Rashba SOC. At lower $\tilde{T}$ and immiscible conditions $\eta_g > 1$, this phase organizes into a stripe phase with quasi-long range orientational and translational order, but this symmetry is spontaneously broken at a critical temperature $\tilde{T}_c \approx 7$ \cite{McGarrigleMicro2023}. This system has more degenerate single-particle states than particles, and as a result there is no singly condensed momentum state. Instead, a spin-correlated, structured normal fluid phase forms, characterized by a circular set of occupied momentum modes.

%\hrulefill
\PRLsep
\vspace{-0.5em}

\setlength{\bibitemsep}{0em} % Adjust the spacing (default is usually larger)
\printbibliography[heading=none]

\end{document}